# Distributed opto-mechanical analysis of liquids outside standard fibers coated with polyimide

HILEL HAGAI DIAMANDI, YOSEF LONDON, GIL BASHAN, AND AVI ZADOK*

Faculty of Engineering and Institute of Nano-Technology and Advanced Materials, Bar-Ilan University, Ramat-Gan 5290002, Israel

**Abstract**

The analysis of surrounding media has been a long-standing challenge of optical fiber sensors. Measurements are difficult due to the confinement of light to the inner core of standard fibers. Over the last two years, new sensor concepts have enabled the analysis of liquids outside the cladding boundary, where light does not reach. Sensing is based on opto-mechanical, forward stimulated Brillouin scattering interactions between guided light and sound waves. In most previous works, however, the protective polymer coating of the fiber had to be removed first. In this work, we report the opto-mechanical analysis of liquids outside commercially available, standard single-mode fibers with polyimide coating. The polyimide layer provides mechanical protection but can also transmit acoustic waves from the fiber cladding towards outside media. Comprehensive analysis of opto-mechanical coupling in coated fibers that are immersed in liquid is provided. The model shows that forward stimulated Brillouin scattering spectra in coated fibers are more complex than those of bare fibers, and strongly depend on the exact coating diameter and the choice of acoustic mode. Nevertheless, sensing outside coated fibers is demonstrated experimentally. Integrated measurements over 100 meters of fiber clearly distinguish between air, ethanol and water outside polyimide coating. Measured spectra are in close quantitative agreement with the analytic predictions. Further, distributed opto-mechanical time-domain reflectometry mapping of water and ethanol outside coated fiber is reported, with a spatial resolution of 100 meters. The results represent a large step towards practical opto-mechanical fiber sensors.

---

* Corresponding author: Avinoam.Zadok@biu.ac.il

**Main Text**

**1. Introduction**

Standard optical fibers are designed to guide a single optical mode at the inner core, with every effort made to prevent propagating light from leaking outside. The same fibers also support a broad variety of guided acoustic modes [1]. Coupling between co-propagating guided light and sound waves in fibers is known for over 30 years [2-5]: Optical fields may stimulate acoustic waves oscillations through electrostriction [2-5], and the resulting strain, in turn, may modulate and scatter guided light due to photo-elasticity [2-5]. Such processes are often referred to as forward stimulated Brillouin scattering (F-SBS). They are studied extensively in standard fibers [2-7], photonic-crystal and micro-structured fibers [8-19], tapered micro-wires [20,21] and even photonic integrated circuits [22-24].

Unlike the optical mode, the transverse profiles of guided acoustic modes span the entire cross-section of the fiber cladding and reach its outer boundary. The oscillations of the stimulated acoustic waves are therefore affected by media outside the fiber. This property is at the basis of a new class of fiber-optic sensors, first proposed by Antman and coworkers in 2016 [25]. The decay rates of stimulated acoustic waves were shown to increase with the mechanical impedance of surrounding liquids [25]. Based on acoustic decay rates, the impedances of water and ethanol could be measured with 1% accuracy [25]. Several other successful demonstrations of this sensing principle quickly followed [26-28]. The first reports were restricted to integrated measurements over the entire lengths of fibers under test [25-28]. However, the concept was recently extended to the distributed analysis of liquids as well [29,30], with up to 3 kilometers range in one report [29] or 15 meters resolution in another [30]. Distributed opto-mechanical measurements may find many field applications in the oil and gas sector, oceanography, desalination processes, chemical industries, and more.

Most measurements reported to-date suffered from one major drawback: the protective polymer coating of the fiber under test had to be completely or partially removed [25-30]. The standard, dual-layer acrylate coating of optical fibers absorbs the acoustic waves that are stimulated in the silica cladding and keeps them from reaching the medium under test. On the other hand, mechanical protection is essential for handling kilometers of fiber. Fortunately, the opto-mechanical sensing of liquids can be performed using commercially available, standard single-mode fibers that are coated by polyimide layers. The polyimide coating is originally designed to protect the fiber at higher temperatures than those sustained by acrylate polymers [31]. In addition, however, the thin polyimide layer also supports the propagation of stimulated acoustic waves towards surrounding media.

Recently, Chow and Thevenaz achieved accurate, integrated F-SBS measurements in 20 meters of polyimide-coated fiber [30,32]. The 80 microns-diameter fiber was immersed in ethanol and water [30,32]. Measurements were supported by approximate analysis based on the analogy between the fiber coating and planar thin films [32]. The model gives useful intuition for coating design, but cannot provide complete predictions of F-SBS spectra. We reported the experimental distinction between air and water in distributed opto-mechanical sensing outside a polyimide-coated fiber [29], without analysis.

In this work, the model of F-SBS is revisited and significantly extended to address coated fibers. The resonance frequencies, decay rates, spatial profiles and magnitudes of opto-mechanical coupling due to radial acoustic modes are calculated below for coated fibers in different liquids, starting from first principles. The analysis shows that F-SBS in coated fibers is more complex than the corresponding process in bare fiber. In particular, the effect of a given test liquid on the F-SBS decay rate may vary between modes and depends strongly on the exact diameter of the coating layer. The analysis is detailed in Section 2. The model provides guidelines for opto-mechanical sensing of test liquids outside polyimide-coated fibers.

Section 3 presents experimental results of integrated opto-mechanical sensing outside 100 meters-long sections of polyimide-coated, standard 125 microns fiber. Measurements clearly distinguish between air, ethanol and water outside the coating. The experimental F-SBS spectra are in very good quantitative agreement with calculations. Lastly, Section 4 shows distributed analysis along 1.6 km of coated fiber. Measurements successfully distinguish between sections of fiber that are immersed in water and ethanol. A concluding discussion is given in Section 5. Preliminary results were presented in recent conferences [33-35].

## 2. Analysis of forward stimulated Brillouin scattering in coated fibers surrounded by liquid

Consider a single-mode optical fiber with silica cladding of outer radius $a$. The fiber is coated with a uniform, radially-symmetric layer of polymer with an outer radius $b$, and immersed in an infinite liquid under test. The coated fiber supports several classes of guided acoustic modes. In this work we focus on radial modes $R_{0,m}$, where $m$ is an integer [2-5]. In these modes, the material displacement vector is radially-symmetric. Each radial mode is characterized by a cut-off frequency [2-5]. Opto-mechanical sensing relies on F-SBS processes that are mediated by these modes [25-30]. Efficient F-SBS requires that the axial phase velocity of the acoustic mode matches that of the guided optical mode [2-5]. This condition is met when the frequency of the acoustic oscillations $\Omega$ is very close to the modal cut-off [2-5]. At that limit, the acoustic wave-vector is nearly entirely radial, and the axial wavenumber of the acoustic wave is negligible [1-5]. The axial component of the material displacement is very small as well, and the displacement vector may be approximated as purely radial [1-5].

Let us denote the velocities of longitudinal acoustic waves as $V_{L,i}$, where $i = 1, 2, 3$ correspond to the silica cladding, the coating layer and the surrounding liquid, respectively. Let $p_i \equiv \Omega/V_{L,i}$ represent the radial acoustic wavenumbers in the three media. Subject to the conditions above, the radial profile $U^{(m)}(r)$ of the material displacement is given by the following expressions [36-39]:

$$\begin{aligned} p_1 A_3^{(m)} J_1(p_1 r) &\equiv U_1^{(m)}(r) & : r \leq a \\ p_2 \left[ A_1^{(m)} J_1(p_2 r) + A_2^{(m)} Y_1(p_2 r) \right] &\equiv U_2^{(m)}(r) & : a \leq r \leq b \\ p_3 A_4^{(m)} \left[ J_1(p_3 r) + j Y_1(p_3 r) \right] &\equiv U_3^{(m)}(r) & b \leq r \end{aligned} \qquad (1)$$

Here $r$ denotes the radial coordinate, $J_1, Y_1$ are the first-order Bessel functions of the first and second kinds, respectively, and $A_{1,2,3,4}^{(m)}$ are constant values, in units of m². The Bessel functions of the second kind are singular at the origin, hence $Y_1$ cannot describe displacement within the silica region. The combination of Bessel functions in the coating layer represents the sum of waves propagating inward and outward along the radial direction. The form $J_1 + jY_1$ of displacement in the liquid region, also known as the Hankel function of the first kind, corresponds to a wave propagating in the outward direction only [36-38].

The radial component of the stress tensor $T_{rr,1}^{(m)}(r)$ in the cladding layer $r \leq a$ is given by:

$$T_{rr,1}^{(m)}(r) = c_{44,1} A_3^{(m)} \left[ (p_1 \kappa_1)^2 J_0(p_1 r) - \frac{2}{r} p_1 J_1(p_1 r) \right] \qquad (2)$$

The corresponding stress element $T_{rr,2}^{(m)}(r)$ within the coating, $a \leq r \leq b$, equals:

$$T_{rr,2}^{(m)}(r) = c_{44,2}(p_2\kappa_2)^2 \left[ A_1^{(m)} J_0(p_2 r) + A_2^{(m)} Y_0(p_2 r) \right]$$
$$-c_{44,2}\frac{2}{r} p_2 \left[ A_1^{(m)} J_1(p_2 r) + A_2^{(m)} Y_1(p_2 r) \right] \qquad (3)$$

In Equations (2) and (3), $J_0$ and $Y_0$ are the zero-order Bessel functions of the first and second kinds, respectively, $c_{44,i}$ denote elements of the stiffness tensors $\mathbf{C}_i$ of media $i = 1, 2$, and $\kappa_i^2 \equiv c_{11,i}/c_{44,i}$ [36-38].

The boundary conditions for guided acoustic waves require that $U_1^{(m)}(a) = U_2^{(m)}(a)$, $U_2^{(m)}(b) = U_3^{(m)}(b)$, $T_{rr,1}^{(m)}(a) = T_{rr,2}^{(m)}(a)$, and that the radial stress at $r = b$ equals the liquid pressure, with an opposite sign [36-38]:

$$T_{rr,2}^{(m)}(b) = \Omega \rho_3 V_{L,3} \frac{J_0(p_3 b) + j Y_0(p_3 b)}{J_1(p_3 b) + j Y_1(p_3 b)} U_3^{(m)}(b) \qquad (4)$$

Here $\rho_3$ is the density of the surrounding liquid. The mechanical impedance of the liquid is given by: $Z_3 = \rho_3 V_{L,3}$ [1].

The set of boundary conditions equations was solved numerically. The explicit form of the coefficients matrix used in subsequent calculations is given in the Supplementary Information. Non-trivial solutions only exist for a discrete set of $\Omega$ values. Each solution defines the coefficients $A_{1,2,3,4}^{(m)}$ (except for a common complex magnitude), and thereby the radial displacement profile of a mode $R_{0,m}$ through Eq. (1).

The elastic parameters of the silica cladding are well known and defined: $a = 62.5$ μm, $c_{44,1} = 31.26$ GPa, $\kappa_1 = 1.586$, and $V_{L,1} = 5969$ m/s [40]. In contrast, the corresponding parameters in polyimide vary among different sources [41-43]. While values of density and the ratio $\kappa_2$ are generally consistent, a wide range of reports is given for Young's modulus [41-43] (and hence for $c_{44,2}$ and $V_{L,2}$ as well). We experimentally estimated the speed of sound in the polyimide coating layer of the specific fiber under test (see in Section 3). The calibrated parameters used in simulations were $b$ = 70.5 µm, $c_{44,2}$ = 2.8 GPa, $\kappa_2$ = 2.031, and $V_{L,2}$ = 2850 m/s. Air ($\rho_3 \sim 0$), ethanol ($\rho_3$ = 789 kg/m$^3$, $V_{L,3}$ = 1143 m/s) or water ($\rho_3$ = 1000 kg/m$^3$, $V_{L,3}$ = 1484 m/s) were considered as media outside the coating.

With air outside the fiber, the frequencies that satisfy the boundary conditions equations are real-valued, denoted hereunder as $\Omega_m$. Each frequency represents the cut-off of a mode $R_{0,m}$. The real-valued frequencies signify that no acoustic dissipation takes place at the boundary between the coated fiber and air. Note that non-zero losses do occur due to bulk acoustic dissipation in both silica and coating, which is not included in the model. Acoustic loss parameters in the coating are unknown. The effect of internal acoustic losses on F-SBS processes was therefore characterized in experiment, as discussed later. Figure 1(a) shows the calculated, normalized transverse profile of mode $R_{0,5}$ with the coated fiber in air. The cut-off frequency $\Omega_5$ is near 2π×178 MHz.

When the surrounding liquid has non-zero mechanical impedance, the boundary conditions are satisfied by complex-valued frequencies: $\Omega_m + j\Gamma_m$. The imaginary component of the frequency represents the decay rate of the acoustic oscillations, due to partial transmission of energy into the liquid. The values of $\Gamma_m$ vary with the liquid impedance $Z_3$. Hence measurements of acoustic decay rates through F-SBS processes are central to opto-mechanical fiber sensing [25-30].

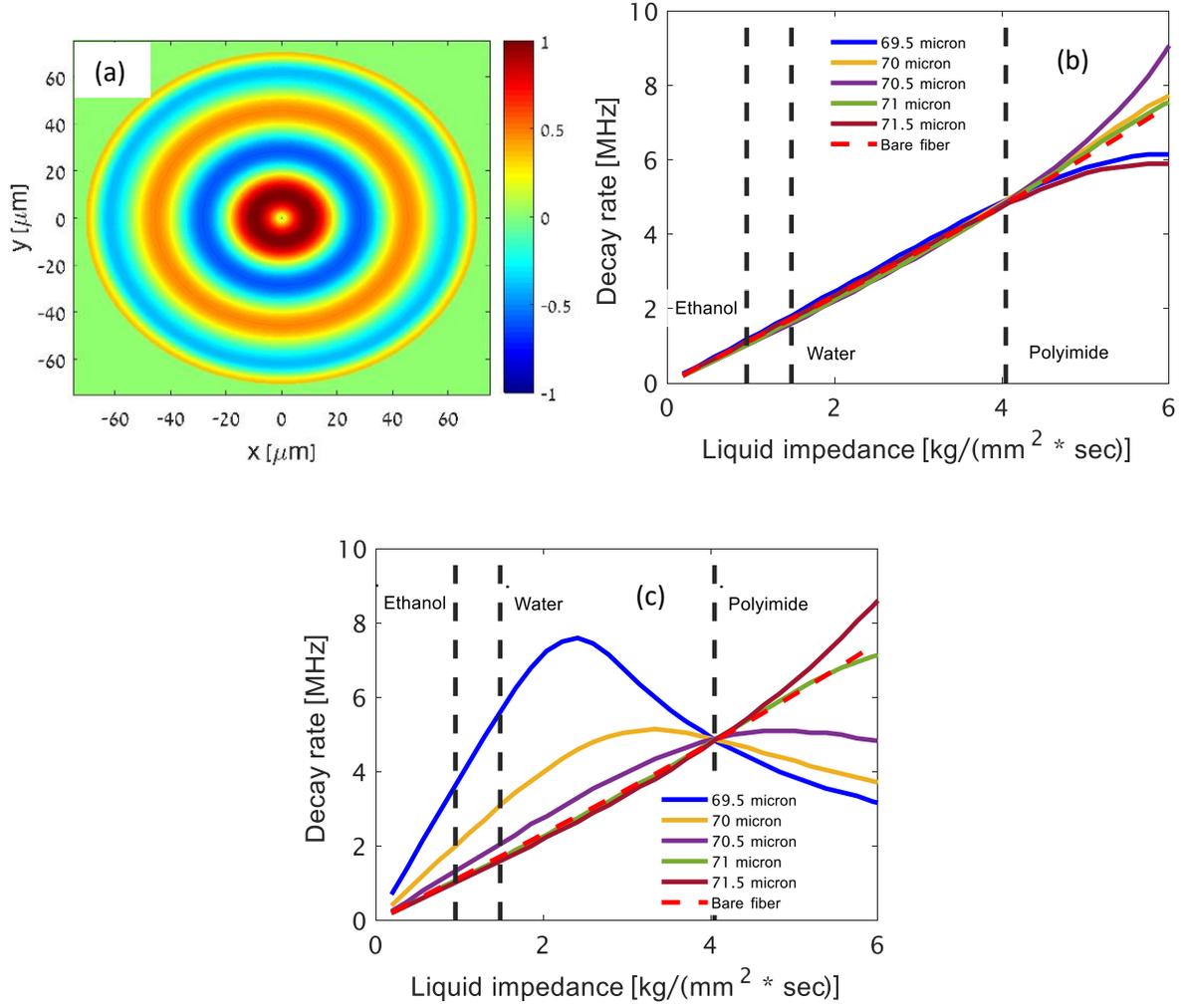

Fig. 1. (a) Solution of guided acoustic modes. Calculated transverse profile of the material displacement in the radial acoustic mode $R_{0,5}$ of a standard fiber, with polyimide coating. The coating radius is 70.5-µm and the fiber is in air. The cut-off frequency of that mode is close to 2π×178 MHz. (b) Calculated decay rates $\Gamma_5/(2\pi)$ of acoustic mode $R_{0,5}$ for different coating radii, as functions of the liquid impedance $Z_3$ (solid traces, see legend). Vertical black lines denote the mechanical impedances of ethanol, water and polyimide. When $Z_3$ is lower than the impedance of the polyimide coating, the decay rates increase monotonously with $Z_3$ and follow closely the analytic predictions for a bare fiber (Eq. (5), red dashed trace). The decay rates exhibit comparatively weak sensitivity to changes in the coating radius. (c) Same as panel (b), for acoustic mode $R_{0,9}$ (cut-off frequency near 2π×325 MHz). The decay rates $\Gamma_9/(2\pi)$ are different from those of mode $R_{0,5}$, and strongly depend on the exact coating radius.

The simulation is validated by taking the limit $b \rightarrow a$. The calculated decay rates in this case converge to those of the acoustic modes in bare fiber [25]:

$$\Gamma_m = -\frac{V_{L,1}}{4a}\ln\left(\left|\frac{Z_3 - Z_1}{Z_3 + Z_1}\right|\right) \quad (5)$$

Here $Z_1$ = 13.13 kg/(mm²×s) is the impedance of silica [40]. The decay rate in bare fiber is a monotonously increasing function of $Z_3$ within the range of interest, it is the same for all radial mode orders $m$, and depends rather weakly on micron-scale changes in $a$. In contrast, the acoustic decay rate for a coated fiber in a given liquid varies with both the coating radius and choice of acoustic mode, as discussed next.

Fig. 1(b) shows the numerically-calculated decay rates $\Gamma_5$ for mode $R_{0,5}$ in the coated fiber, as functions of $Z_3$. Calculations are provided for different coating radii $b$. When the mechanical impedance of the liquid is lower than that of polyimide ($Z_2$ = 4.04 kg/(mm²×s)), the modal decay rates follow closely the bare fiber solution, and show comparatively modest sensitivity to changes in $b$. The calculated modal decay rates in ethanol and water are 2π×1.1 MHz and 2π×1.7 MHz, respectively. These rates change by ±7% with ±1 µm variations in coating radius.

Figure 1(c) presents the corresponding calculations for mode $R_{0,9}$ (cut-off frequency $\Omega_9$ close to 2π×325 MHz). The decay rate $\Gamma_9$ of that mode is strongly dependent on sub-micron variations in the coating radius. The calculated values with the fiber in ethanol or in water may change by 250% when the radius is modified by just ±1 µm. Such changes are well within the specified tolerance for the coating thickness. The decay rate $\Gamma_9$ may be very different from that of a bare fiber, or that of mode $R_{0,5}$. The decay rates for all modes and radii converge to that of a bare fiber when the liquid impedance $Z_3$ matches $Z_2$, as might be expected. When $Z_3$ is larger than the coating impedance $Z_2$, decay rates vary with diameter and choice of modes even more intensely (see for example in panel (b)). Due to the comparative robustness of $\Gamma_5$ to uncertainty in coating radius $b$, the distributed analysis reported in Section 4 was carried out using mode $R_{0,5}$.

The geometric sensitivity of the guided acoustic modes is due to interference effects within the coating layer. The radial periodicities of the displacement profiles in polyimide are only few microns, on the same scale as the uncertainty in coating thickness. Figure 2(a) shows calculations of $U^{(9)}(r)$ for two coating radii: 69.5 µm and 70.5 µm, with the fiber in air. The two displacement profiles within the coating layer are markedly different. Figure 2(b) presents the corresponding calculations for the displacement profile $U^{(5)}(r)$. The change in radius induces smaller modifications to the radial profile of mode $R_{0,5}$. The decay rates of higher-order modes tend to be more sensitive to geometric variations, due to the shorter radial periods of their displacement profiles. In addition, geometric sensitivity of the decay rates increases with modal confinement in the polyimide layer (see also Supplementary Information).

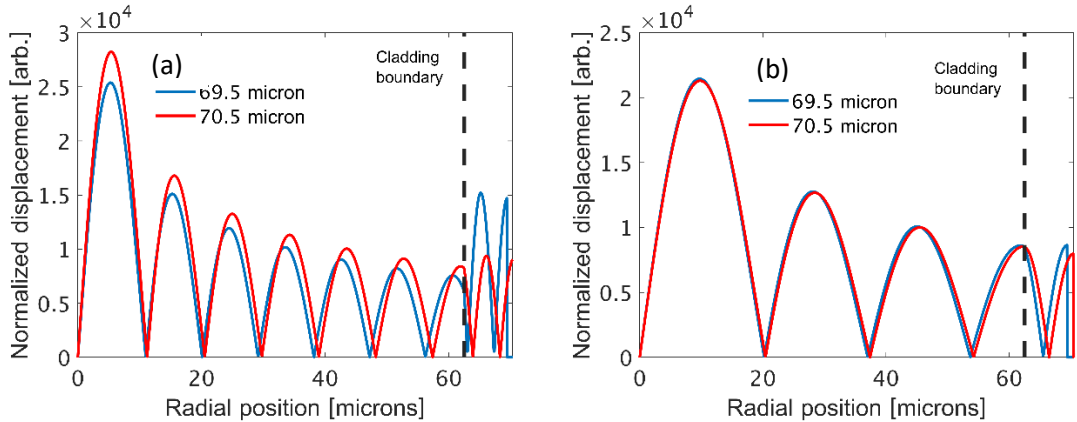

Fig. 2. (a) Calculated normalized radial profiles of material displacement of mode $R_{0,9}$ within the silica and polyimide layers, when the fiber is kept in air. The coating outer radius is 69.5 µm (blue), or 70.5 µm (red). The vertical black line denotes the boundary between cladding and coating. (b) Same as panel (a), for mode $R_{0,5}$.

Unlike the radial modes of a bare fiber [25], there is no single value for the acoustic decay rate for a coated fiber in a given liquid. The calculations suggest that the decay rates are difficult to predict a-priori. Possible ellipticity and non-concentricity of the coating layer, and variability in the elastic parameters of polyimide, may add to the uncertainty of modelling. The decay rates in a specific sensing fiber may require pre-calibration in a set of known test liquids. Additional analysis is provided in the Supplementary Information.

Consider next an optical pump wave in an F-SBS process, with instantaneous power $P(t)$ where $t$ stands for time. We denote the radio-frequency Fourier transform of $P(t)$ as $\tilde{P}(\Omega)$. The pump wave may stimulate the oscillations of the guided radial acoustic modes through electrostriction [2-5,25,39]. The acoustic waves then induce photo-elastic scattering of a co-propagating optical probe wave at a different optical wavelength. The photo-elastic scattering, in turn, manifests in cross-phase modulation of the probe wave at the output of a fiber of length $L$ [2-5,25,39]. The magnitude of the probe wave phase modulation through mode $R_{0,m}$ may be expressed as [25,39]:

$$\delta\tilde{\phi}_m(\Omega) = \gamma_{0,m}(\Omega) L \tilde{P}(\Omega) \qquad (6)$$

In Eq. (6), $\gamma_{0,m}(\Omega)$ represents the opto-mechanical nonlinear coefficient of F-SBS due to mode $R_{0,m}$, in units of (W×m)$^{-1}$ [17,39]. The opto-mechanical coefficient is mode and frequency dependent. It is given by [17,39]:

$$\gamma_{0,m}(\Omega) \equiv \frac{k_0}{16 n^2 c \rho_1} \frac{Q_{ES}^{(m)} Q_{PE}^{(m)}}{\Omega_m \Gamma_m} \frac{1}{1 - j(\Omega - \Omega_m)/\Gamma_m} \qquad (7)$$

Here $\rho_1$ and $n$ are the density and refractive index of silica respectively, $c$ is the speed of light in vacuum, $k_0$ is the vacuum wavenumber of the probe wave, and $\Omega_m$ notes again the cut-off frequency of radial guided acoustic mode $R_{0,m}$. $Q_{ES}^{(m)}$ denotes the overlap integral between the transverse profile of the electrostrictive driving force and that of the acoustic displacement $U^{(m)}(r)$ [39]. Lastly, $Q_{PE}^{(m)}$ is the overlap integral between the transverse profile of the strain associated with $U^{(m)}(r)$ and that of the optical mode [39]. The instantaneous phase modulation of the probe wave $\delta\phi(t)$ may be calculated by taking the inverse-Fourier transform of $\delta\tilde{\phi}_m(\Omega)$, summed over all modes: $\delta\phi(t) = \sum_m \int \delta\tilde{\phi}_m(\Omega) \exp(j\Omega t) d\Omega$.

The nonlinear F-SBS coefficients of specific modes are of Lorentzian line-shapes, with widths that are given by $\Gamma_m$ [25]. Note that the widths are narrower than the separation between the resonance frequencies of adjacent modes, so that F-SBS at a given acoustic frequency takes place through a single radial mode, at most. Measurements of the F-SBS spectra $\gamma_{0,m}(\Omega)$ may retrieve $\Gamma_m$, and thereby provide estimates of $Z_3$ [25-30]. Figure 3 shows the calculated nonlinear coefficients as a function of frequency, for a fiber with 70.5 µm-radius polyimide coating in ethanol and water. According to the simulation, the magnitude of $\gamma_{0,m}$ at the strongest resonance peaks reaches the order of 2-2.5 (W×km)$^{-1}$. The spectral linewidths in ethanol are consistently narrower than those in water. The analysis suggests that measurements of F-SBS spectra may distinguish between the two liquids outside the coating. This prediction is tested experimentally in the next sections.

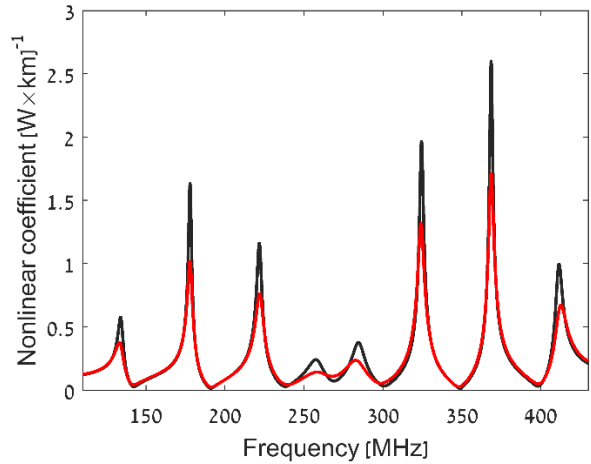

Fig. 3. Calculated magnitude of the nonlinear coefficient of forward stimulated Brillouin scattering in a standard fiber, with polyimide coating of 70.5 µm radius. The fiber is immersed in ethanol (black) or water (red).

## 3. Integrated sensing outside coated fiber

Integrated opto-mechanical sensing outside polyimide-coated fibers was performed first. Experiments were based on F-SBS phase modulation of an optical probe wave, as described in Eq. (6). Due to the phase-matching characteristics of F-SBS, the phase modulation of the probe is non-reciprocal: a probe wave that is co-propagating with pump pulses is modulated due to F-SBS, whereas a counter-propagating probe wave is almost unaffected. This property forms the basis of the measurement principle [9,25,39,44,45].

The experimental setup was first proposed by Kang and coworkers [9], and was used by us in previous works [25,37,42,43]. It is illustrated in Fig. 4 [25]. Light from a first distributed-feedback laser diode at 1559 nm wavelength was the source of the optical pump wave. Pump light first passed through a semiconductor optical amplifier (SOA). The drive current of the SOA was modulated by repeating pulses of 4 ns duration and 10 µs period. The SOA provided the necessary high extinction ratio of over 30 dB, however it cannot support pulse durations below few ns without significant distortion. Therefore, pulses were then further modulated in an electro-optic amplitude modulator (EOM) connected in series, which was driven by 1 ns-long pulses with the same period. The pulse generators used for the SOA and EOM were synchronized. Lastly, pump pulses were amplified by an erbium-doped fiber amplifier (EDFA) to an average power of 100 mW and launched into a section of fiber under test. The pump pulses stimulated guided radial acoustic modes of the fiber. The polarization of the pump pulses was scrambled to suppress the stimulation of torsional-radial guided acoustic modes [26,28,46].

The fiber under test was placed within a Sagnac interferometer loop ([9,25,39,44,45], Fig. 4). Pump pulses propagated along the fiber in the clockwise direction only, and were blocked by a tunable optical bandpass filter (BPF) from reaching the loop output. Continuous probe light from a second distributed laser feedback diode at 1550 nm wavelength was coupled into the loop in both directions. The BPFs were tuned to transmit the probe wave. As noted earlier, only the clockwise-propagating probe was subject to phase-matched F-SBS modulation $\delta\phi(t)$. For parameters relevant to this work, $\delta\phi(t)$ is much smaller than $2\pi$. The non-reciprocal phase modulation was converted to intensity modulation of the probe wave at the loop output [9,25,39,44,45]. A polarization controller within the loop was used to bias the relative phase between clockwise and counter-clockwise propagating probe waves [25,39,44,45]. The output probe was detected by a photo-receiver. The receiver voltage $V(t)$ was proportional to $\delta\phi(t)$ [25,44,45]. The proportionality constant depended on the optical power of the probe wave, the responsivity of the detector, and the states of polarization along the setup. The receiver voltage was sampled by a digitizing oscilloscope for further offline processing. Traces were averaged over 4,096 repeating pulses.

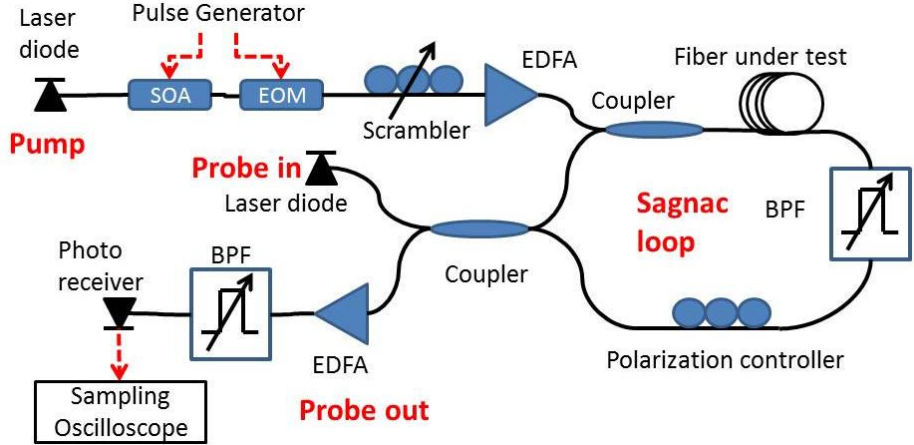

Fig. 4. Experimental setup for integrated opto-mechanical sensing outside polyimide-coated fibers under test [25]. SOA: semiconductor optical amplifiers. EOM: electro-optic amplitude modulator. EDFA: erbium-doped fiber amplifier. BPF: tunable optical bandpass filter

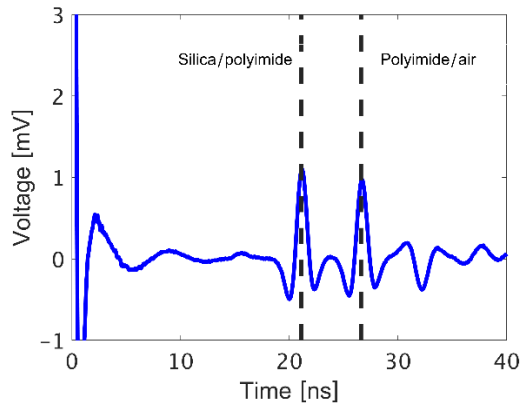

Fig. 5. Measured voltage of the detected output probe wave as a function of time. Vertical dashed lines denote the phase modulation of the probe wave by two acoustic impulses. The first corresponds to acoustic reflection between cladding and coating, and the second is due to reflection at the outer coating boundary.

Three sections of fibers from the same manufacturer were tested. The coating radii of all samples were specified as 70±2 µm, and their lengths were between 50-100 meters. Measurements were taken with the fibers in air, ethanol and water. Figure 5 presents an example of the detected output probe trace $V(t)$, for a fiber in air. A large impulse at $t = 0$ is due to nonlinear-index cross-phase modulation of the probe by the intense pump pulse [9]. This event provides a timing reference. The two subsequent, weaker impulses correspond to acoustic reflections: first at the boundary between silica and polyimide, and then between polyimide and air. The 5.5 ns time delay between the two reflections corresponds to the two-way acoustic time-of-flight across the coating. This delay provided an estimate for the acoustic velocity in the polyimide coating: 2850 m/s. This speed corresponds to Young's modulus of 7.5 GPa, a value which is within the range of literature reports [41-43].

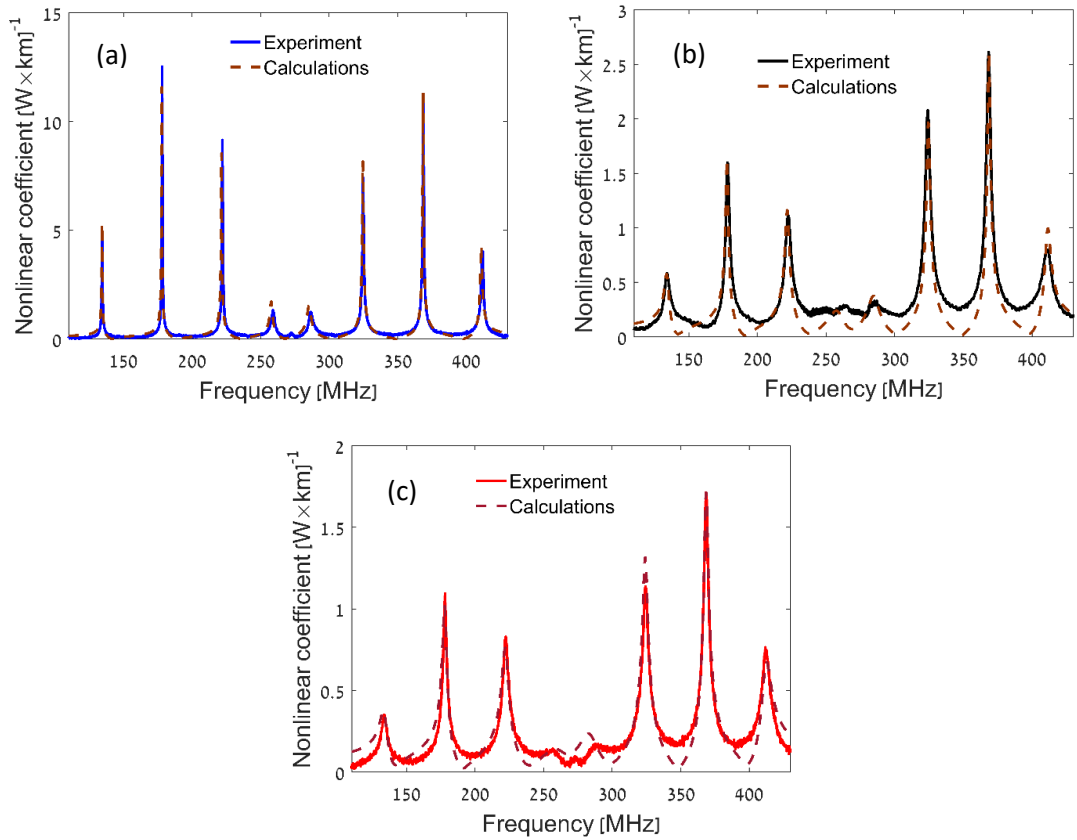

Fig. 6. Integrated measurements (solid traces) and calculations (dashed traces) of forward stimulated Brillouin scattering spectra in a polyimide-coated fiber. The coating radius was specified as 70±2 µm and fitted as 70.5 µm. (a) Fiber in air. (b) Fiber in ethanol. (c) Fiber in water. Very good agreement is obtained between analysis and experiment.

Figure 6 shows several measurements of the ratio between the Fourier transform of the output voltage $|\tilde{V}(\Omega)|$, and that of the pump pulses $|\tilde{P}(\Omega)|$ (solid traces). Calculated F-SBS spectra are plotted as well (dashed traces). The measured spectra were first normalized to a maximum value of unity. Next, each spectrum was further scaled to align its maximum value with the largest

calculated nonlinear coefficient in the same test medium $|\gamma_{0,m}(\Omega)|$, in units of [W×km]$^{-1}$. Note that the analysis of liquids does not rely on the absolute magnitudes of nonlinear coefficients, but rather on the linewidths of their spectra. The linewidths are unaffected by constant scaling.

Panel (a) presents the results of calculation and measurement for the fiber in air. The F-SBS spectrum consists of sharp and narrow resonances. The acoustic decay rates are slow, as expected, due to negligible dissipation at the outer boundary of the coating. The measured $\Gamma_5$ of mode $R_{0,5}$ in air was 2π×(0.13±0.02) MHz. The modal linewidths are determined by acoustic losses within the cladding and coating layers. Since bulk acoustic losses in polyimide were not known, the experimental linewidths of the F-SBS resonances were used to define the values of $\Gamma_m$ in the simulations. The observed resonance frequencies and relative strengths of F-SBS interactions through the different radial modes are well accounted for by the analysis. The best fitting between measured and calculated spectra was achieved for a coating radius of 70.5 µm.

Panels (b) and (c) show the measured and calculated F-SBS spectra of the fiber in ethanol and water, respectively. Compared with the fiber in air, the measured F-SBS spectra in liquids are characterized by weaker and broader peaks, in agreement with expectations. The acoustic decay rates due to outward acoustic radiation at the boundary were calculated directly as described in Section 2, with no fitting parameters. The decay rates due to internal losses, as measured in Fig. 6(a), were added to the calculated values of $\Gamma_m$. The measured F-SBS spectra for both liquids are in quantitative agreement with corresponding calculations.

Lastly, Fig. 7 compares between integrated measurements of the F-SBS spectra in water and ethanol. The two spectra are distinct. The modal decay rates in ethanol are consistently slower than in water, as anticipated. The measured rates for mode $R_{0,5}$ were 2π×(1.1±0.05) MHz and 2π×(1.5±0.25) MHz in ethanol and water, respectively. The experimental errors represent variations between the measurements of three samples. The larger variations among decay rates in water may have to do with the adsorption of water in polyimide (see also discussion in Section 5). The rates are in very good agreement with the analytic predictions of 2π×1.2 MHz in ethanol and 2π×1.8 MHz in water (with internal losses included). Mode $R_{0,5}$ was used in distributed sensing of the two liquids, as discussed in the next section.

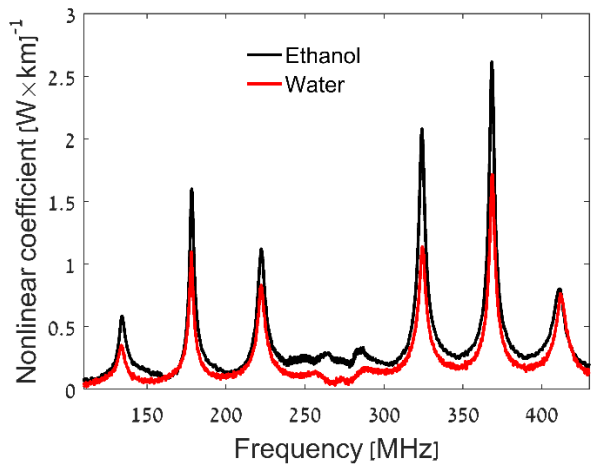

Fig. 7. Integrated measurements of forward stimulated Brillouin scattering spectra over 100 meters of a polyimide-coated fiber under test. The coating radius is specified as 70±2 µm. The fiber was immersed in ethanol (black) or in water (red). Each trace was multiplied by a different constant scaling factor, so that the highest values of the measured nonlinear coefficients matched the results of calculations.

## 4. Distributed sensing outside coated fiber

The scaling of the opto-mechanical sensing concept from integrated to distributed analysis has presented a fundamental challenge. All known protocols for distributed fiber sensing, using Rayleigh [47], Raman [48], or Brillouin scattering [49,50], are based on backwards-scattering, whereas guided acoustic modes scatter light in the forward direction. Backscatter events can be directly localized using time-of-flight measurements, however forward scattering may not. Very recently, our group [29] and the Thevenaz group at EPFL, Switzerland [30] proposed and demonstrated two strategies to work around this difficulty. Distributed analysis of F-SBS has been reached using both.

Here we apply the opto-mechanical time-domain reflectometry scheme of [29] towards the analysis of liquids outside polyimide-coated fiber. The measurement principle is reiterated here only briefly. The reader is referred to [29] for full detail. Consider two continuous optical fields of frequencies $\omega_{1,2} = \omega_0 \pm \frac{1}{2}\Omega$, where $\omega_0$ is a central optical frequency. The two tones co-propagate along a fiber under test and may stimulate a guided acoustic mode. The stimulation of the acoustic wave is associated with the coupling of optical power, from the higher-frequency optical field to the lower-frequency one. The evolution of the local power levels $P_{1,2}(z)$ of the two fields, where $z$ is axial position along the fiber, is given by the following pair of coupled nonlinear differential equations:

$$\frac{dP_{1,2}(z)}{dz} = -\alpha P_{1,2}(z) \mp \gamma_{0,m}(\Omega, z) P_1(z) P_2(z) \tag{8}$$

Here $\alpha$ is the coefficient of linear propagation losses along the fiber. Subtraction of the two equations yields:

$$\gamma_{0,m}(\Omega, z) = \frac{1}{2P_1(z)P_2(z)} \frac{d[P_2(z) - P_1(z)]}{dz} + \frac{1}{2P_1(z)P_2(z)} \alpha [P_2(z) - P_1(z)] \tag{9}$$

Equation (9) suggests that measurements of the two local power levels may retrieve local opto-mechanical coupling spectra. The experimental procedure relies on time-domain reflectometry to obtain estimates of $P_{1,2}(z)$ [29]. To that end, the amplitudes of the two optical field components are jointly modulated by repeating, isolated pulses. Rayleigh backscatter contributions at frequencies $\omega_{1,2}$ are separated by a narrowband Brillouin fiber amplifier, detected, sampled and processed according to Eq. (9). Measurements are repeated for multiple values of frequency offset $\Omega$, to obtain maps of F-SBS as a function of position and frequency. Noise associated with the Rayleigh backscatter of coherent light is reduced by the averaging of each collected trace over many choices of the central optical frequency $\omega_0$, while keeping $\Omega$ the same. Additional noise mitigation is achieved by reducing the coherence of the input optical signals through random phase modulation [29,51].

The experimental setup is shown in Fig. 8. It is based on that of [29] with several changes. Light from a tunable laser diode at optical frequency $\omega_0$ was used as the source of all optical waveforms. The laser output passed through an electro-optic phase modulator, which was driven by a repeating pseudo-random binary bit sequence at 75 Mbit/s rate from a pattern generator. The output voltage of the generator matched $V_\pi$ of the modulator. Phase modulation broadened the laser linewidth, and partially suppressed measurement noise due to coherent Rayleigh backscatter [51].

The modulated waveform was split in two paths. Light in the sensor branch (Fig. 8) was amplitude-modulated in a double-sideband EOM. The modulator was biased for carrier suppression and driven by the output voltage of a sine-wave generator of variable radio-frequency $\frac{1}{2}\Omega$. The modulated waveform consisted of two tones, at optical frequencies $\omega_{1,2} = \omega_0 \pm \frac{1}{2}\Omega$. The optical field was then amplified by an EDFA, and amplitude-modulated by pulses of 1 µs duration and 25 µs period in a second EOM. The peak power of the pulses was 250 mW. The pulse duration corresponds to a spatial resolution of 100 m. The dual-tone pulses were then routed into the fiber under test through a fiber-optic circulator.

Rayleigh backscatter at frequencies $\omega_{1,2}$ propagated through the circulator into a 150 meters-long section of standard fiber which served as a frequency-selective Brillouin amplifier. The Brillouin pump wave was generated in the amplifier branch of the laser diode output (Fig. 8). Light was upshifted in frequency in a single-sideband electro-optic modulator (SSB), which was driven by a sine wave from the output of a microwave generator. The offset frequency $\Omega_{pump}$ was chosen as either $\Omega_B + \tfrac{1}{2}\Omega$ or $\Omega_B - \tfrac{1}{2}\Omega$, where $\Omega_B = 2\pi \times 10.73$ GHz is the Brillouin frequency shift in the Brillouin amplifier. The pump wave was then amplified by second EDFA to an output power of 250 mW and launched into the opposite end of Brillouin amplifier segment through a second circulator. The former choice of $\Omega_{pump}$ provided selective amplification of Rayleigh backscatter from the fiber under test at frequency $\omega_1$, and the latter choice led to Brillouin gain at $\omega_2$.

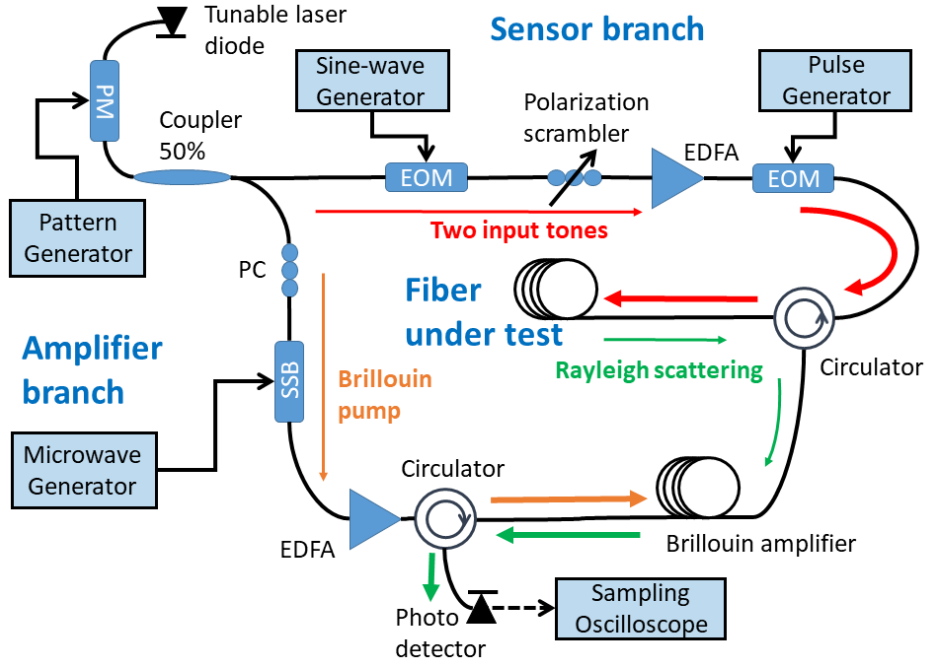

Fig. 8. Schematic illustration of the experimental setup used in opto-mechanical time-domain reflectometry of a polyimide-coated fiber. EOM: electro-optic amplitude modulator. SSB: Single-sideband electro-optic modulator. PM: electro-optic phase modulator. EDFA: erbium-doped fiber amplifier. PC: polarization controller [29].

Rayleigh backscatter traces at the output of the Brillouin amplifier were detected by a photo-receiver of 200 MHz bandwidth, and sampled by a digitizing oscilloscope at 5 ns intervals. Measurements were averaged over 64 repeating input pulses, and digitally filtered by a moving-average window of 750 ns duration. Data was collected for 1,024 choices of central optical frequency $\omega_0$ for each choice of frequency offset $\Omega$, within a wavelengths range of 1559±0.5 nm. The entire experimental procedure was repeated for 40 choices of $\Omega$ near $\Omega_5$ of the fiber under test.

Figure 9(a) shows an opto-mechanical time-domain reflectometry map of a 1.6 km-long section of the same polyimide-coated fiber that was measured in Section 3 above. The entire length of the fiber under test was immersed in water, except for a 200 meters-long section located 1.3 km from the input end. That section was placed in ethanol instead. The section located in ethanol is characterized by a stronger F-SBS peak, with a narrower linewidth.

Figure 9(b) shows the measured linewidths of mode $R_{0,5}$ as a function of position. The linewidth is $2\pi \times (1.35 \pm 0.1)$ MHz within the section that was placed in ethanol, and $2\pi \times (2.2 \pm 0.2)$ MHz everywhere else. Note that the F-SBS spectra are broadened by the finite duration of optical pulses [29]. The linewidth in ethanol is narrower than in water, in agreement with the results of simulations and integrated sensing. The distributed analysis successfully distinguishes between the two liquids, and correctly locates the section immersed in ethanol. The linewidth in water is broader than those of integrated measurements (Fig. 7). The difference may have to do with softening of the coating, following the adsorption of water in the polyimide. Due to the long acquisition duration of opto-mechanical time-domain reflectometry, the fiber was kept in water for many hours. The coating does not adsorb ethanol. Figure 9(b) also shows the local resonance frequency $\Omega_5(z)$ (bottom trace). Unlike the modal linewidth, the F-SBS resonance is unaffected by the surrounding liquid.

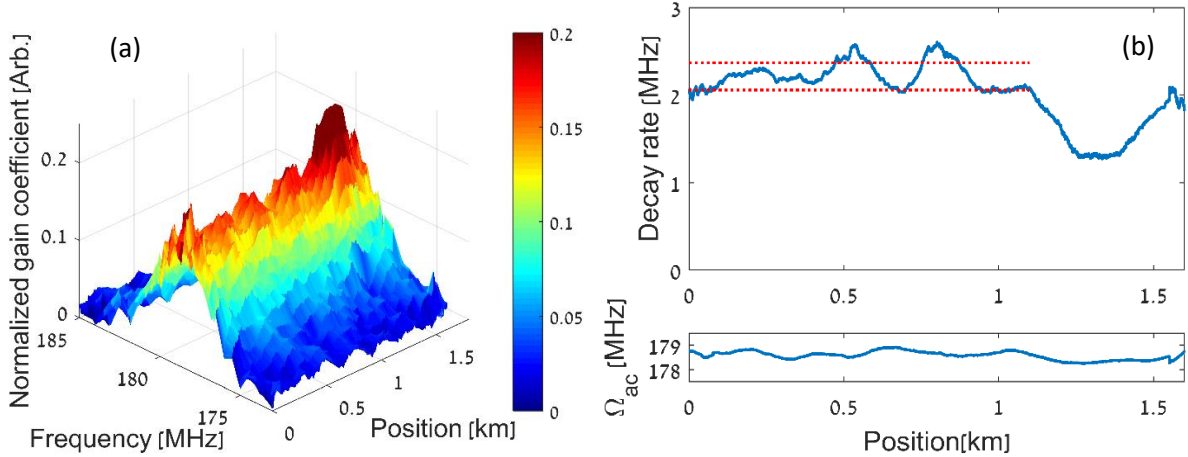

Fig. 9. Opto-mechanical time-domain reflectometry measurements of a coated fiber. (a) Map of the normalized opto-mechanical gain coefficient $|\gamma_{0,5}(\Omega)|$ of mode $R_{0,5}$, as a function of frequency and position along 1.6 km of fiber under test. The fiber is coated with polyimide layer of 70±2 µm outer radius and was kept in water for most of its length. A 200 meters-long section located 1.3 km from the optical input end was immersed in ethanol instead. (b) Measured F-SBS linewidth $\Gamma_5/(2\pi)$ as a function of position (upper trace). The section of fiber placed in ethanol is characterized by a slower decay rate, in agreement with simulations and integrated sensing measurements. Horizontal red lines denote the range of experimental uncertainty in the measurement of the decay rate in water. The lower trace shows the measured resonance frequency $\Omega_5/(2\pi)$.

## 5. Summary and discussion

The analysis of F-SBS processes promises to open new horizons for fiber-optic sensors, extending beyond where guided light may reach directly. The recent introduction of protocols for distributed F-SBS analysis raises interest even further [29,30]. However, the sensing of media outside optical fibers may be practical only if fibers remained protected. The inner layer of standard, dual-acrylate coating is very soft, and strongly attenuates elastic perturbations. Therefore, it prevents the acoustic waves that are stimulated within the cladding from reaching a substance under test. In this work, we examined opto-mechanical sensing over fibers with polymer coating. We find that unlike the standard dual-layer acrylate coating, the stiffer polyimide coating layer provides sufficient transmission of the acoustic waves to distinguish between surrounding liquids.

The analytic model for F-SBS was extended to include the coating layer. Formalism was developed for calculating the cut-off frequencies and transverse profiles of radial acoustic modes that are guided by the silica-polyimide structure. The model also evaluates the strengths of F-SBS interactions between guided light and each radial acoustic mode. Analysis is in good agreement with integrated measurements of F-SBS spectra over sections of coated fibers in air. The relative magnitudes and resonance frequencies of the F-SBS peaks are accounted for by calculations.

When the coated fiber is placed in a liquid medium of non-zero mechanical impedance, the acoustic oscillations decay due to the radiation of energy from the coating outward. Simulations predict the decay rates, and the F-SBS spectra, as functions of coating radius, liquid impedance, and choice of mode. Unlike the case of a bare fiber, the F-SBS spectra show large sensitivity to small-scale changes in coating diameter and large variations among modes. Nevertheless, the acoustic decay rates for a given liquid impedance may still be predicted, at least for some acoustic modes. The more exact predictions of the analysis complement the useful and intuitive simplified model provided recently in [32].

In the experimental part of this work, we calibrated the acoustic velocity and the radius of the polyimide layer using integrated analysis of F-SBS over 100 meters-long sections of fiber [25]. Measured F-SBS spectra with the fiber in air, ethanol and water were in very good agreement with calculations. The measurements could distinguish between the two liquids outside the fiber. Finally, we performed opto-mechanical time-domain reflectometry of the coated fiber in liquids [29]. The obtained map of F-SBS coefficients resolved ethanol and water outside the coating, and properly identified the location of a section of fiber immersed in ethanol.

Significant challenges remain for ongoing and future work. Specific coating materials may be developed for the purpose of opto-mechanical sensing. Much like impedance matching in radio-frequency and optical waveguides, the coating material can be designed for optimum response to a range of liquid impedance values of interest. Better control over the geometry and thickness of the coating might be necessary to that end. Many potential applications require the additional protection of coated fibers within cables. Sensing of liquids with cabled fibers is still a formidable task.

Opto-mechanical sensing can benefit from effects that the test liquid itself may have on the coating material. For example, the absorption of water in polyimide leads to swelling of the coating layer, and the application of strain on the fiber [52]. The effect on the coating is reversible. This property is at the basis of point-sensors of relative humidity, using fiber Bragg gratings that are recoated with polyimide [52]. F-SBS analysis of coated fibers can therefore enable indirect distributed sensing of relative humidity. The principle can be extended to specific distributed detection of other chemical reagents, based on their effect on properly functionalized coating layers. On the other hand, prolonged interaction with the substance under test may modify the pre-calibration of the F-SBS sensor or degrade the long-term reliability of the coating. The analytic and experimental protocols proposed in this work may also serve for characterization and quality control of the coating layers themselves. This prospect was already demonstrated above, in the estimate of the effective Young's modulus for radial displacement in the polyimide layer.

In conclusion, this work provides proof of concept for distributed opto-mechanical analysis of media outside commercially available coated fibers. The results mark a major milestone for this new approach to optical fiber sensing. Future work would look to improve the range, resolution and accuracy of the measurements, and explore potential applications.

**Acknowledgement**

The authors acknowledge the financial support of the European Research Council (ERC), grant no. H2020-ERC-2015-STG 679228 (L-SID); and of the Israeli Ministry of Science and Technology, grant no. 61047. H. Hagai Diamandi is grateful to the Azrieli Foundation for the award of an Azrieli Fellowship.